\documentclass[10pt,journal]{IEEEtran}
\usepackage{setspace}
\usepackage{cite}
\usepackage{graphicx,balance}

\usepackage[cmex10]{amsmath}
\usepackage{amsfonts}
\usepackage{subcaption}
\usepackage{longtable,balance}
\usepackage{array}
\usepackage{multirow}
\usepackage{url}
\usepackage{color}
\usepackage{threeparttable}
\usepackage{rotating}

\makeatletter

\newcommand{\Rmnum}[1]{\expandafter\@slowromancap\romannumeral #1@}
\makeatother
\usepackage{algorithmic}
\usepackage{algorithm}
\allowdisplaybreaks[4]
\begin{document}
\title{From Resource Auction to Service Auction: An Auction Paradigm Shift in Wireless Networks}

\author{Xianhao Chen,
        Yiqin Deng,
        Guangyu Zhu,
        Danxin Wang,
        and~Yuguang Fang,~\IEEEmembership{Fellow,~IEEE}% <-this % stops a space
\thanks{Xianhao Chen, Guangyu Zhu, and Yuguang Fang are with the Department of Electrical and Computer Engineering, University of Florida, Gainesville, FL 32611, USA. (e-mail: xianhaochen@ufl.edu, gzhu@ufl.edu, fang@ece.ufl.edu).}% <-this % stops a space
\thanks{Yiqin Deng is with the School of Computer Science and Engineering, Central South University, Changsha 410083, China (e-mail: dengyiqin@csu.edu.cn).}
\thanks{Danxin Wang is School of Computer Science and Hubei LuoJia Laboratory, Wuhan University, Wuhan 430072, China,  (e-mail: wangdanxin@whu.edu.cn;).}
\thanks{This work was supported in part by US National Science Foundation under grants CNS-2106589 and IIS-1722791. Corresponding author: Yiqin Deng.}}
\maketitle
\begin{abstract}
In 5G and beyond, the newly emerging services, such as edge computing/intelligence services, may demand the provision of heterogeneous communications, computing, and storage (CCS) resources on and across network entities multi-hop apart. In such cases, traditional resource-oriented auction schemes, where buyers place bids on resources, may not be effective in providing end-to-end (E2E) quality-of-service (QoS) guarantees. To overcome these limitations, in this article, we coin the concept of E2E service auction where the auction commodities are E2E services rather than certain resource. Under this framework, buyers simply bid for services with E2E QoS requirements without having to know the inner working (which resources are behind). To guarantee E2E QoS for winning bids while ensuring essential economic properties, E2E service auction requires addressing the joint problem of network optimization and auction design with both economical and QoS constraints. To substantiate the mechanism design, we illustrate how to devise E2E service auctions for edge computing systems under various scenarios. We also identify the research opportunities on E2E service auction mechanism design for other critical use cases, including edge intelligence.
\end{abstract}

\IEEEpeerreviewmaketitle
\section{Introduction\label{sect: introduction}}
Recent years have witnessed numerous transformative applications, such as virtual/augmented reality, video analytics, autonomous driving, and smart healthcare~\cite{mao2017survey}, which would make our lives more connected and ``smarter''. To support these applications, staggering amount of data must be collected and transported to desired locations for consumption, storage, and/or processing for intelligence extraction. For this reason, 5G and beyond (5G+) is designed to be a concerted supporting framework for communications, computing, and storage (CCS), rather than a framework for data communications only. For the effective provision of 5G+ services, the joint management of CCS resources is the key.

Unfortunately, no matter how much CCS resources are provisioned in a 5G+ system, due to the tremendous interests in emerging applications, the system operator always faces resource shortage and has to outsource external CCS resources to serve its customers' demands. Auction is typically an effective way to stimulate the resource sharing for such purpose. Thus, auction mechanism design has gained tremendous attention from the networking community over the past decades. In wireless networks, traditional auction mechanisms typically ask buyers to place bids on resources (i.e., computing resource at a server~\cite{sun2018double} or spectrum over a region~\cite{zhou2009trust,chen2019privacy}). This paradigm, called ``resource auction'' in this article, enables dynamic resource sharing among multiple parties over wireless networks.

Resource auctions mostly focus on the trading of either communications or computing resource alone. However, to accommodate 5G+ systems, auction mechanisms should be designed in the way that CCS resources are provisioned and traded as a whole. Consequently, traditional resource auctions face two major challenges. First, since they generally focus on either communication or computing alone, they may not be effective in providing E2E QoS guarantees for many emerging 5G+ services. For example, to deliver data stream from end devices to edge nodes for computing, an edge computing service may demand a combination of heterogeneous CCS resources on and across network entities multi-hop apart. In such cases, obtaining sufficient spectrum or computing resource alone does not provide any guarantee for E2E latency. The service latency could be intolerable due to network traffic congestion, even though there is significantly powerful computing resource.

Second, resource auctions may incur excessive complexity on the user side. In resource auctions, end users have to learn the network environments, like the availability and quality of spectrum bands~\cite{li2019service}, the capabilities of in-situ computing, and the distances to computing servers~\cite{sun2018double}, in order to valuate these networking/computing resources and place bids accordingly. Letting end users select and valuate resources to support services on their own, could dramatically increase the operational overheads on buyer side, especially when an E2E service involves multitype resources. Typically, a buyer only knows the desired service and the QoS requirement, while viewing the service's internal implementation as a ``blackbox''. To implement and commercialize auction mechanisms over wireless networks, it is crucial to make auction process more user-friendly.

Based on the above observations, in this article, we advocate the paradigm design shift from resource auction to ``service auction'' for wireless networks. Specifically, we propose a general service auction framework, called \textit{E2E service auction} framework, referring to the auction scheme that commodities are E2E services. The word ``end-to-end'' underscores that the auction provides everything (resources) needed for winning buyers from the beginning (when users triggers) to the end (when it gets the results). As its name suggests, there are two salient features of E2E service auction:

\begin{itemize}
\item E2E service auction supplies all needed CCS resources to guarantee the E2E QoS for service buyers. In other words, the auction scheme not only provides computing and storage resource (if needed for the considered service type), but also support E2E data transmissions between sources and destinations one-hop or multi-hop apart.

\item Buyers initiate service requests with QoS requirements without having to know the inner working (which CCS resources are behind).
\end{itemize}

The first feature provides E2E QoS guarantees, while the second feature makes the auction process more user-friendly compared with resource auctions where users have to valuate and bid for resources.

We remark that E2E service auction is different from combinatorial auctions, where each bidder bids for a bundle of resources~\cite{zhang2012auction}. Unlike E2E service auction where buyers simply submit service requests with QoS requirements, in a combinatorial auction, a buyer needs to select and valuate bundles of resources, say CCS resources, properly. For end users with limited knowledge of the network, this is not an easy task to accomplish and not user-friendly. Additionally, combinatorial auction could induce excessive computational complexity for large instances, as the winner determination problem for combinatorial auction is generally NP-hard~\cite{zhang2012auction}.

In the context of edge computing, our recent work \cite{chen2021end} designs a double-sided E2E service auction for edge computing systems. However, the idea of E2E service auctions has not yet been cultivated as a broadly applicable design principle, which we believe, could inspire more further interesting research. For this reason, unlike \cite{chen2021end} that addresses one concrete double-sided computing market, this article attempts to present the general architecture and design principle for E2E service auctions, expand its application to multiple edge computing/intelligence scenarios, and discuss the future research directions.

The remainder of this article is organized as follows. Section \ref{sect: Approaches} describes the basics of auction mechanisms and other economic approaches. Section \ref{sect: system} introduces the general E2E service auction model. Section \ref{sect: usecase} illustrates how to design E2E service auction for edge computing services under several concrete scenarios. Section \ref{sect: opportunities} identifies the research opportunities for E2E service auctions. Section \ref{sect: conclusion} concludes the article.

\section{Economic Approaches for Wireless Networks\label{sect: Approaches}}
To enable effective resource sharing and pricing over wireless networks, there are several widely-used economic approaches. Before elaborating on E2E service auction, we first give a brief overview of these market mechanisms.

\subsubsection{Pricing\label{subsect: Pricing}}
Directly setting prices is a very common market mechanism. Pricing schemes can be broadly classified into static pricing and dynamic pricing, depending on whether it can accommodate varying network conditions and demands. Game theory is a powerful tool for pricing. However, appropriate pricing relies on the sufficient knowledge of the values of resources and/or the valuations of agents, which may be hard to obtain.

\subsubsection{Contract Theory\label{subsect: Contract}}
Contract theory is effective when the information about users is incomplete. In such a case, a service provider offers a contract and then each end user/resource supplier chooses the best contract items to maximize its utility. Nevertheless, contract theory still requires certain information (probability distributions) about agents. When the network scale is small, contract theory is not be effective enough, because the historical statistics (probability distributions) may not reflect the real-time demands and supplies well.

\subsubsection{Auction Approaches\label{subsect: Auction}}
Auction is suitable for a networking market where a service provider has incomplete information about agents, or even has no prior information at all. This is because truthful auction can elicit the valuation information from buyers and/or sellers through the bidding process. Due to this salient advantage, auction approaches have been extensively exploited for resource allocation and incentive design for wireless networks.

An auction market contains buyers, sellers, and an auctioneer. Buyers and sellers submit bid and ask prices to auctioneer, respectively, for certain commodities. Then, the auctioneer determines winning buyers and sellers as well as the clearing prices for both sides~\cite{zhang2012auction}. Based on the competition behaviors among agents, we can classify auction approaches into forward auction, reverse auction, and double auction.

\textbf{Forward auction:} Multiple buyers bid for the commodities offered by a single seller. In wireless networks, forward auction generally addresses the allocation of resources to end users.

\textbf{Reverse auction:} Multiple sellers compete to sell their commodities to a single buyer. In wireless networks, reverse auction can be employed to create incentives for resource suppliers.

\textbf{Double auction:} Multiple buyers and multiple sellers co-exist in a market. By introducing the competitions to both sides, double auction not only incentivizes sellers to share resources, but also effectively allocates resources to end users.

\section{The Paradigm Shift to E2E Service Auction\label{sect: system}}
In this section, we present a general E2E service auction framework, and discuss the potential auction approaches and design requirements.

\subsection{System Architecture\label{subsect: Architecture}}
\begin{figure}[t]
\centering
\includegraphics[width=3in]{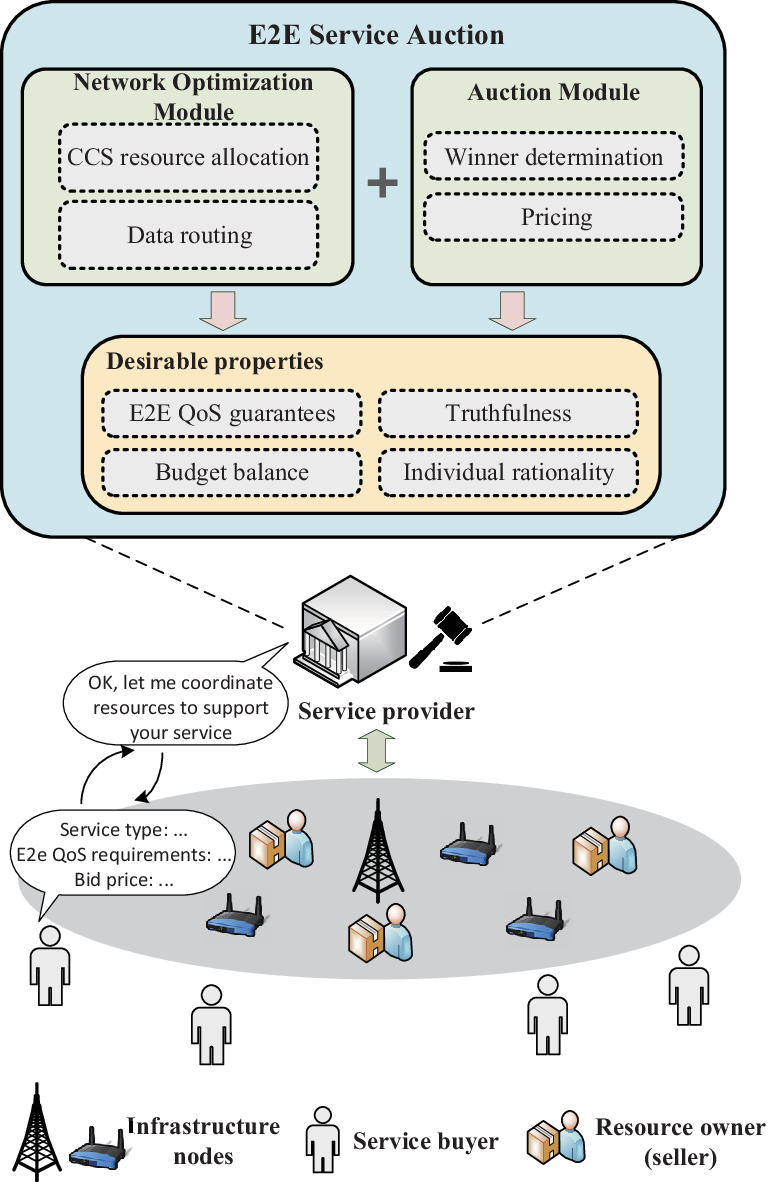}
\caption{A general E2E service auction framework. For generality, we do not specify which wireless services the buyers request, and which kinds of resources the sellers own. As shown in the dialogue, a buyer only needs to claim his/her service request with E2E QoS requirements and a bid price, while the service provider takes charge of coordinating appropriate resources and data routing to satisfy the buyer's needs.}
\label{fig: SA_framework}
\end{figure}

A general service auction market is illustrated in Figure \ref{fig: SA_framework}. In general, there are multiple buyers (end users), and one or multiple sellers (resource owners). To sell services with QoS guarantees, a certain level of centralized control is necessary. The service provider serves as the central entity providing services to the buyers, and also acts as the auctioneer in the market\footnote{We assume that the service provider (auctioneer) is trustworthy, i.e., executing the auction mechanism faithfully. This is reasonable in practice, as a wireless service provider has the motivation to maintain its reputation.}. The service provider can be a broker who harvests spare resources and infrastructure from sellers to provide services. It can also be a traditional cellular service provider which attempts to enhance the system capacity based on auction approaches. To conduct auction and network optimization, the service provider should have some basic spectrum bands for gathering information and exchanging control signaling messages by following the design principle of software-defined networking (SDN).

As discussed earlier, buyers in a service auction market only need to know service types and QoS requirements, such as data rate and/or latency requirements, for their applications. Each buyer submits a service request with E2E QoS requirements and a bid price, as shown in the dialogue in Figure \ref{fig: SA_framework}. If there are resource owners in addition to the service provider, such as selfish spectrum holders and server owners, these resource owners act as sellers, placing ask prices for supplying resource. After gathering the bid and network information, the service provider determines the winners and pricing, and allocates all the needed resources, potentially purchased from selfish resource owners, to support the services for winning buyers with the required QoS.

The service provider has two modules: auction module and network optimization module. Auction module takes charge of winner determination and pricing, while ensuring some essential economic properties, like truthfulness, individual rationality, and budget balance (which will be introduced in Section \ref{subsect: properties}). Unlike resource auctions, E2E service auction features a network optimization module that manages heterogeneous CCS resources (e.g., transmit powers, spectrum bands, and virtual machines/containers) and data routing, to support the winning services with E2E QoS guarantees. In other words, E2E service auction design requires addressing the joint problem of network optimization and auction design. In a nutshell, the main design challenge for E2E service auctions is twofold:

\begin{itemize}
\item E2E service auction might involve the provisioning of CCS resources together, rather than communication or computing resource alone as usually done in typical auctions in wireless networks.

\item E2E service auction requires solving a joint network optimization (including resource management and data routing) and auction design problem. This is fundamentally different from conventional auction mechanisms where winner determination and pricing are the only outputs. In addition to winner determination and pricing, E2E service auction should coordinate network-wide CCS resources and data routing under E2E QoS constraints.
\end{itemize}

\subsection{Design Requirements\label{subsect: properties}}
In general, E2E service auction should preserve the following desirable properties.

\textbf{Truthfulness:} No buyer/seller can improve his/her utility by claiming a bid/ask price deviating from the true valuation/cost. Truthfulness reduces the cost of auction by eliminating bidders' incentives to spend resources on learning others' strategies and determining the optimal bidding strategy.

\textbf{Individual Rationality:} No buyer pays more than his/her bid price, and no seller is paid less than his/her ask price.

\textbf{Budget balance:} The auctioneer gains a non-negative revenue.

\textbf{Computational efficiency:} The auction mechanism should be computationally efficient.

\textbf{E2E QoS Guarantees:} The E2E QoS requirements from winning service buyers should be satisfied.

\section{Use Case: Edge Computing\label{sect: usecase}}
\begin{figure*}[t]
\centering
\includegraphics[width=5in]{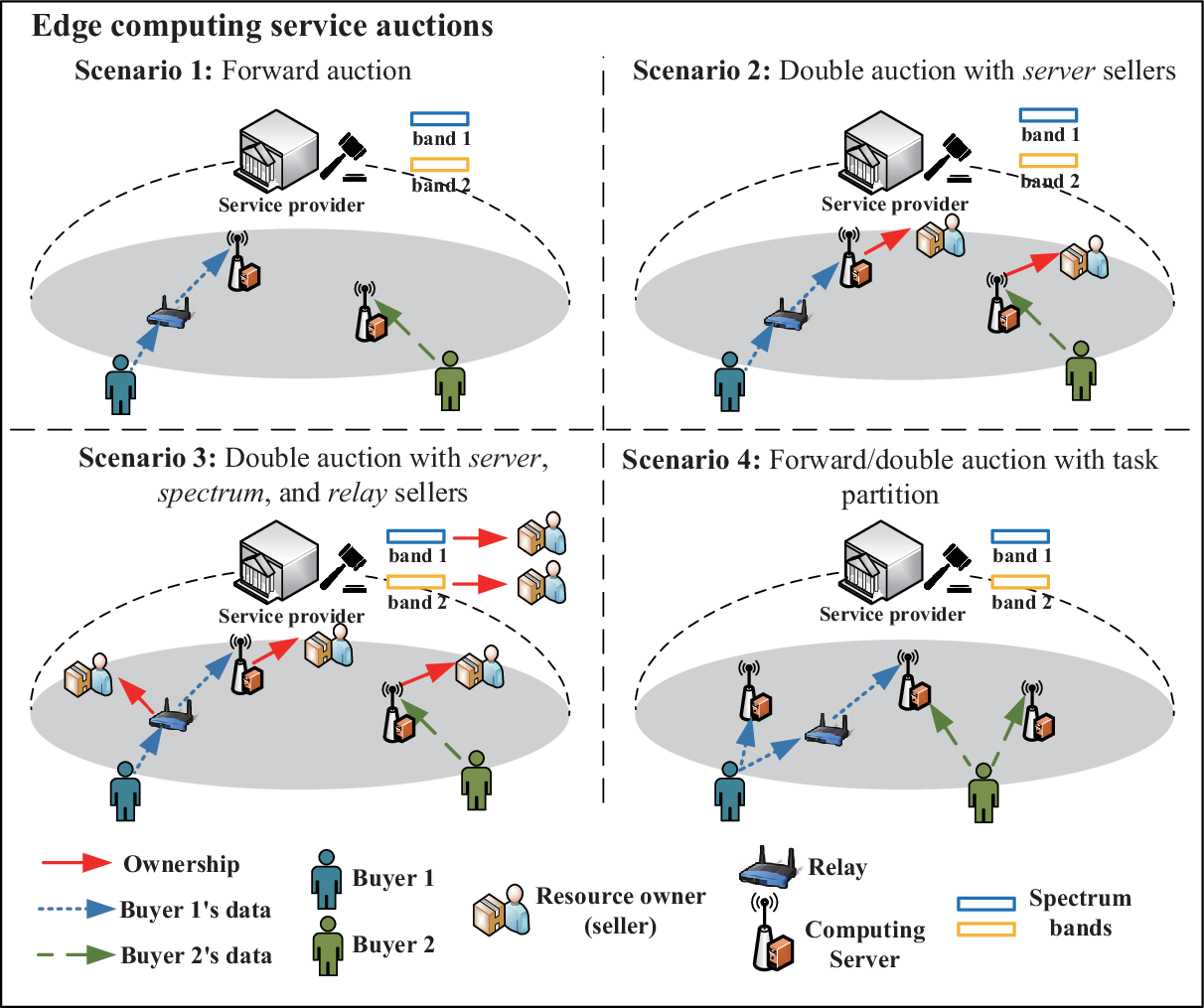}
\caption{Four concrete scenarios for edge computing service auction. Red solid arrows indicate that the resources are owned by selfish sellers, where incentives are needed to stimulate them to share resources.}
\label{fig: SA_scenarios}
\end{figure*}
In this section, to substantiate the design of our service auction framework, we use edge computing to shed light on how to design E2E service auctions. Edge computing requires the holistic design for CCS, because input data should be delivered from end devices to edge servers for processing. Unfortunately, although some auction schemes have been proposed for edge computing systems~\cite{sun2018double,ma2021tcda}, they focus on computing aspect without taking networking aspect (e.g., spectrum allocation and data routing) into account. The only exception is our work \cite{chen2021end}, as alluded in the introduction. In what follows, by generalizing and extending the situation considered in \cite{chen2021end}, we provide four useful scenarios of edge computing service auctions, each requiring specific design and considerations.

\subsection{Service and E2E QoS Model\label{subsect: service}}
To begin with, we introduce the service request model. We assume that there are $I$ buyers in the considered wireless network. Buyer $i$ initiates $K_i$ service requests, and submits bid price $b_{i,k}$ for his/her $k$-th request.

Depending on service type, a buyer can claim E2E QoS requirements consisting of E2E latency, E2E data rate, CPU frequency, memory space, reliability, delay jitter, and so on. In this article, we specifically consider E2E QoS requirements $QoS_{i,k}=(\theta_{i,k}, \delta_{i,k}, r_{i,k})$, where $\theta_{i,k}$ is the computing requirement, $\delta_{i,k}$ is the storage requirement, and $r_{i,k}$ is the E2E data rate requirement. Such QoS requirements are applicable to real-time processing applications with continuous data streams, such as video analytics applications~\cite{ding2018beef}. Note that it is possible to adopt other QoS models. The format of $QoS_{i,k}$ does not change our fundamental design principle.

\subsection{Scenario 1 (Forward Auction)\label{subsect: computingsingle}}
Let us start with the simplest case where a service provider possesses all resources for edge computing service provisioning, including spectrum bands, computing servers, and relays, as shown in Scenario 1 in Figure \ref{fig: SA_scenarios}. The service provider can adopt forward auction to sell services to users. To support the buyers' requests with E2E QoS requirements $QoS_{i,k}$, one needs to solve a joint network optimization and forward auction problem under QoS constraints. One may adapt the well-known Vickrey-Clarke-Groves (VCG) scheme to maximize the social welfare (the total utility of all participants), as done in \cite{li2016users}. Specifically, given concrete network configurations, one can formulate the edge computing service provisioning problem (which, for instance, jointly determines request assignment, computing resource allocation, spectrum allocation, and data routing) with the objective of social welfare maximization. To achieve truthfulness, the clearing price for each winner is set to his/her marginal harm caused to other participants, following the basic idea of VCG auctions.

However, VCG-style auctions are generally computationally intractable, as they rely on the socially optimal allocation. When taking spectrum allocation into account, the network optimization problem in wireless networks is generally NP-hard~\cite{ding2018beef}, thereby hindering the applications of VCG-style auctions to large-sized networks. To resolve this problem, we can resort to greedy allocations. Since greedy allocations often follow monotonic allocation rule, they tend to guarantee the truthfulness for single-parameter systems (where each buyer submits one bid) according to Myerson's characterization. In \cite{li2013designing}, Li et al. design a truthful forward auction mechanism based on a greedy routing and spectrum allocation scheme. Although their scheme is designed for multi-hop data delivery, the basic idea may be applied here. The detailed mechanism design can be left as the future work.

\subsection{Scenario 2 (Double Auction with Server Owners) \cite{chen2021end} \label{subsect: computingdouble}}
A service provider may lack computing resource to accommodate the demands on edge computing services. To harvest spare computing resources, double auction is the most appropriate auction approach, whereby the service provider balances the demands and supplies between buyers (users) and sellers (server owners), as shown in Scenario 2 in Figure \ref{fig: SA_scenarios}.

In addition to the buyer-side model presented in Section \ref{subsect: service}, there can be $J$ sellers offering ask prices for sharing their computing resources. We assume that each seller has limited computing and storage resource, and therefore can only support limited number of requests. Let $a_{j,i,k}$ denote seller $j$'s ask price towards buyer $i$'s $k$-th request, depending on their diverse resource utilization and consumption. To meet the criteria in Section \ref{subsect: properties}, one should address a joint network optimization and double auction problem. Unfortunately, this is a very challenging research task.

To address the above problem, we propose to use a two-step approach to effectively decouple network optimization and double auction. At the first step, we cast and solve a ``pure'' network optimization problem that determines service assignment while jointly allocating communication and computing resources to meet the QoS requirements. This problem does not take economics (bid and ask prices) into consideration. For example, one can maximize the system throughput by solving the following problem

\begin{gather}
\textbf{P1:} \quad \max_{h_{i,k}^j, \boldsymbol x, \boldsymbol f} \sum_{1\leq i\leq I}\sum_{1\leq k\leq K_i}\sum_{1\leq j\leq J} h_{i,k}^j r_{i,k}, \\
\textit{subject to E2E QoS constraints},\notag
\end{gather}
where $\boldsymbol x$ is the collection of spectrum allocation variables, and $\boldsymbol f$ represents the data flow rate over each wireless link. $h_{i,k}^j$ is a binary variable, where $h_{i,k}^j=1$ indicates that buyer $i$'s $k$-th request is assigned to seller $j$'s server, and $h_{i,k}^j=0$ otherwise. The E2E QoS constraints ensure that the communication-computing requirements $QoS_{i,k}=(\theta_{i,k}, \delta_{i,k}, r_{i,k})$ must be satisfied under the limited spectrum resource and servers' capabilities. We leave the QoS constraints unspecific to make the problem generic. We call the set of service assignments (indicated by $h_{i,k}^j=1$) obtained from $\textbf{P1}$ as candidate service assignments.

\begin{figure}[tp]
\centering
\begin{subfigure}[t]{0.8\linewidth}
\centering
\includegraphics[width=2.5in]{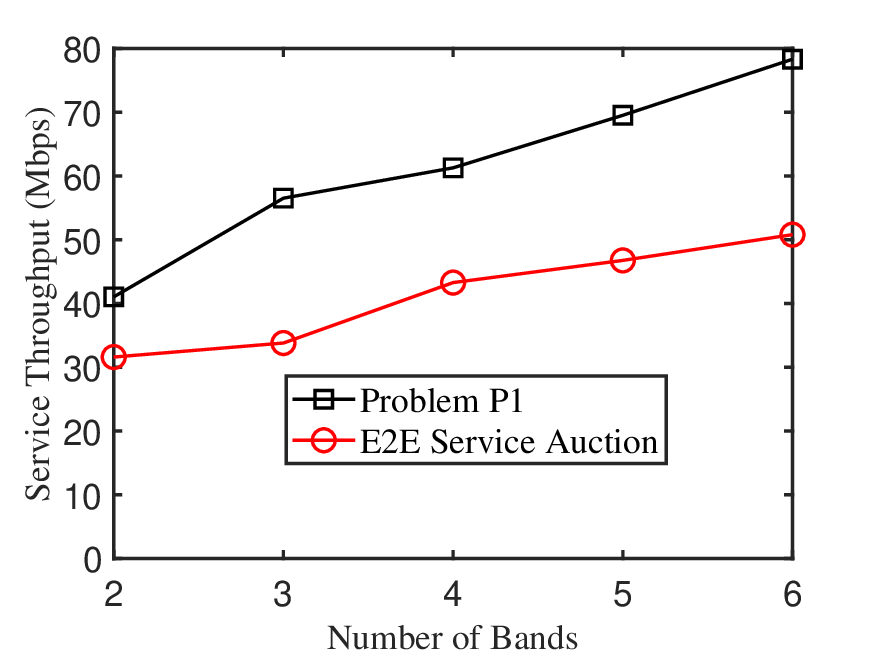}
\caption{The average system throughput versus the number of bands.}
\label{fig: band}
\end{subfigure}

\begin{subfigure}[t]{0.8\linewidth}
\centering
\includegraphics[width=2.5in]{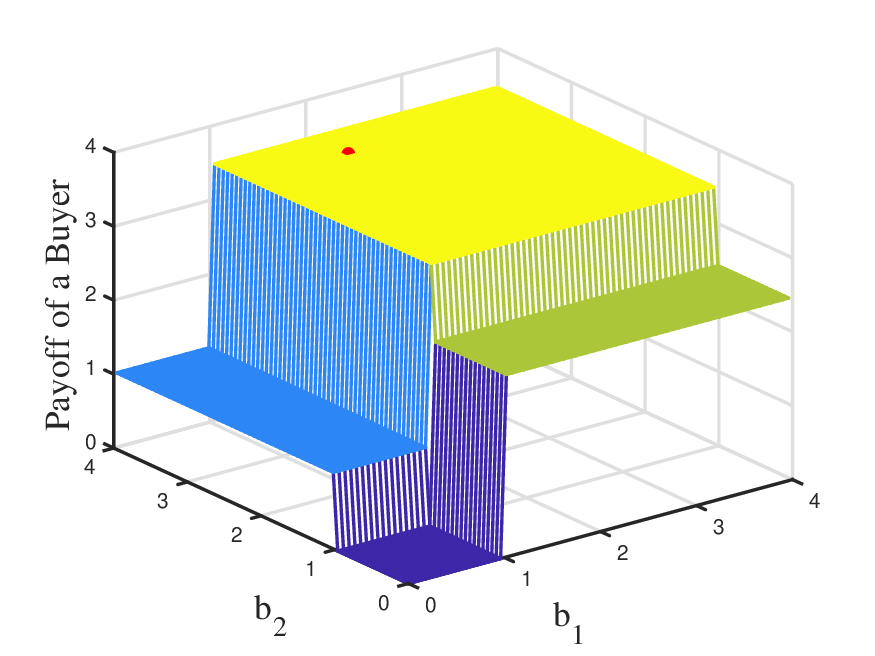}
\caption{The utility of a randomly chosen buyer versus its two bid prices $b_1$ and $b_2$, under the setting of $4$ bands.}
\label{fig: truthfulness}
\end{subfigure}
\caption{Simulation results for the E2E service auction mechanism for Scenario 2 \cite{chen2021end}. We consider a wireless mesh network in a 1000$\times$1000$m^2$ area, with $3$ sellers, 4 relays, and $12$ buyers, each submitting $2$ requests. $r_{i,k}$, $\theta_{i,k}$, $\delta_{i,k}$, $b_{i,k}$, $a_{j,i,k}$ are uniformly drawn from $[4, 8]$Mbps, $[1, 4]$GHz, $[1, 3]$GB, $[0.5, 4]$, and $[0, 1]$. The computing and storage capabilities of servers are uniformly drawn from $[6, 14]$GHz and $[8, 24]$GB, respectively. In the network, each band has the bandwidth of $10$ MHz.}
\label{fig: simulation}
\end{figure}

At the second step, we select winners from the candidate service assignments, and then determine clearing prices for the both sides. Since the set of winning service assignments are a subset of the candidate assignments obtained from $\textbf{P1}$, their QoS requirements $Q_{i,k}$ are certainly satisfied. Moreover, the winner determination and pricing at the second step should guarantee the truthfulness, individual rationality, and budget balance. In this way, E2E service auction provides QoS guarantees while maintaining the desirable economic properties. The interested readers are referred to our work \cite{chen2021end} for the detailed mechanism design.
Figure \ref{fig: band} illustrates the system throughput versus the number of bands over a wireless mesh network. In the figure, ``Problem P1'' is the solution to the ``pure'' network optimization problem $\textbf{P1}$, and ``E2E Service Auction'' is the proposed double auction mechanism. The gap between the two curves is the performance degradation due to the economic impact (i.e., the cost of achieving individual rationality, budget balance, and truthfulness). To demonstrate the truthfulness, Figure \ref{fig: truthfulness} randomly chooses one buyer and evaluates its payoff. The buyer proposes two bid prices for its two service requests. By manipulating its bid prices, the buyer changes its payoff. However, it can be observed that the truthful bidding strategy (the red point) is optimal for the buyer.

\subsection{Scenario 3 (Double Auction with Multitype Resource Owners)}
In addition to computing resource, the service provider may also be short of communication resources. In Scenario 3 in Figure \ref{fig: SA_scenarios}, the service provider lacks computing resource, spectrum resource, and infrastructure nodes (relays), and tends to acquire all of them from the auction market for service provisioning. As a result, server owners, spectrum owners, and relay owners all serve as the sellers for resource sharing.

Designing service auction mechanisms under such scenarios are obviously challenging. To solve the problem, we still formulate a network optimization problem similar to \textbf{P1} to ensure E2E QoS, while allowing more kinds of agents to act as sellers in the double auction. The solution to \textbf{P1} will produce complicated buyer-seller assignments. For example, buyer 1 may be assigned with multiple spectrum owners, multiple relay owners, and one server owners, in the sense that the E2E service may demand multiple bands, multiple relays, and one edge server. The service provider can ask a buyer to submit bid price $b_{i,k}$ together with a vector $\boldsymbol P_{i,k}$, say, $\boldsymbol P_{i,k}=\{0.2, 0.3, 0.5\}$, to indicate the fraction of the bid price given to spectrum owners, relay owners, and server owners, respectively. Once the buyer wins the bid, his/her payment will be shared proportionally among these sellers. The detailed mechanism design will be investigated in the future.

\subsection{Scenario 4 (Forward/Double Auction with Task Partition)}
There are still many variants of mechanism design for edge computing settings. Typically, we assume that that one service request can be at most assigned to one server. However, when parallel execution is feasible, a buyer's input data can be delivered to multiple edge computing servers, each performing a subtask, to speed up the process by making full use of the available computing-communication resources, as illustrated in Scenario 4 in Figure \ref{fig: SA_scenarios}. For instance, it may be viable to partition video data into multiple segments and process them on multiple edge servers in parallel. On the other hand, to improve the service reliability, one task can be duplicated and assigned to multiple sellers. Task duplication is especially useful when edge servers are personal devices and hence are inherently unreliable. In these cases, according to the ownership of the resources, both forward auction and double auction can be adopted to design the service auction mechanism, like Scenario 1-3. If employing double auction, the payment collected from a buyer should be shared among multiple server owners due to the parallel execution. For effective service provisioning, a task must be judiciously partitioned and assigned such that the buyer's QoS requirements can be met and the buyer's payment can compensate the involved resource owners.

\section{Research Opportunities\label{sect: opportunities}}
In addition to edge computing, the design philosophy of E2E service auction can be applied to many other emerging use cases requiring the joint design for CCS. In this section, we discuss the research opportunities on E2E service auction mechanism design.

\subsection{Edge Intelligence Service Auction\label{subsect: intelligence}}
Edge intelligence refers to AI-empowered edge computing. Nowadays tremendous data analytics applications are built on artificial intelligence (AI) algorithms.  When performing model inference, a pre-trained AI model should be sent to and stored at the edge nodes and then input data should be delivered to them for processing or computing. To handle the intensive computing workload of deep neural networks (DNNs), the inference tasks can be partitioned and distributed over multiple devices to enable fast inference, where data shuffling is needed to exchange intermediate results between devices.

Edge nodes cannot only perform inference, but also facilitate model training. When dealing with private-insensitive data, data owners can directly transfer their data and models to edge nodes for training. Otherwise, privacy-preserving learning approaches, such as federated learning and split learning, can come to rescue. In federated learning, data owners train their models on local devices and exchange models with edge nodes for aggregation\cite{ng2020joint}. In split learning, edge nodes partially take over the training load by partitioning a model into two or multiple pieces, thus relieving the computing load on end devices\cite{park2021communication}. In both cases, models and/or intermediate computing results should be frequently exchanged between distributed devices.

For both model inference and training, holistic design for CCS is the key. Under auction framework, trading either communication or computing resource alone is obviously not sufficient to fulfill the tasks with latency guarantee. Spectrum allocation, computing resource allocation, and service request routing should be jointly optimized under QoS constraints. Moreover, the ultimate goal of model inference and training is perfect execution. Therefore, in addition to the QoS metrics in Section \ref{subsect: service}, these learning tasks can also incorporate other novel QoS requirements, such as model loss. In this way, one may reduce communication overhead as long as the semantic representations are still useful for training/inference, following the design philosophy of semantic communications.

\subsection{Mobility-Aware Computing Service Auction\label{subsect: mobility}}
5G+ is expected to support high-quality services for users with high mobility. Imagine that a vehicular user wants to enhance its gaming experience by harnessing the capabilities of edge nodes. Due to the high mobility, not only the channel conditions between end users and base stations vary rapidly, but also the application instances may need to be migrated to new locations closer to users in order to provide satisfactory QoS.

The difficulty in developing mobility-aware service auction comes from the fact that the auction and network optimization may need to be conducted over two different time scales. On the one hand, network optimization must be designed to fit the changing network conditions and user movement, for which the time scale should be short. On the other hand, to reduce signaling overhead and possible service disruption, auction results should be effective during a relatively longer time. Therefore, when performing auction, the service provider should take the movement of end users into account so that CCS resources on users' future trajectories can be reserved.

\subsection{Network Slicing\label{subsect: slicing}}
In our previous mechanisms, end users serve as the buyers for services. Another interesting scenario is that, multiple service providers may act as buyers to purchase infrastructure and/or network resources from infrastructure providers (InPs) to support the service demands from their customers. This process can be enabled by network slicing technology which divides physical network into multiple logical networks (i.e., slices)\cite{habiba2018auction}, creating business opportunities for InPs and service providers by dynamically sharing network resources according to their aggregated E2E QoS demands.

\subsection{Machine Learning based Auction Mechanism Design\label{subsect: ML}}
Recently, deep learning (DL) has been employed to develop truthful auction mechanisms to enhance the economic efficiency\cite{luong2020machine}. E2E service auction addresses the joint problem of network optimization and auction design, which is quite challenging. Therefore, it is hard to provide the exactly or nearly optimal solution when guaranteeing the essential economic properties, particularly truthfulness. To enhance the performance for E2E service auction, applying machine learning tools to E2E service auction is a promising research direction.

\section{Conclusion\label{sect: conclusion}}
In this article, we have argued that the traditional resource auctions for wireless networks might incur user-side overhead and lack QoS guarantees. To remedy these issues, we have advocated the paradigm design shift from resource auction to E2E service auction. Due to the holistic design for auction mechanism and network optimization, the proposed E2E service auction framework guarantees the QoS for winning requests, while ensuring some essential economic properties. We have used several use cases, i.e., edge computing and edge intelligence, to illustrate the design philosophy of E2E service auction. We hope that this article can spark the research interests in E2E service auction.

\bibliography{mybibtex}

\end{document}